# An efficient singlet-triplet spin qubit to fiber interface assisted by a photonic crystal cavity

Kui Wu[1], Sebastian Kindel[2], Thomas Descamps[2], Tobias Hangleiter[2], Jan Christoph Müller[2], Rebecca Rodrigo[1], Florian Merget[1], Hendrik Bluhm[2], and Jeremy Witzens[1]

[1]Institute of Integrated Photonics, RWTH Aachen University, Aachen, 52074, Germany
[2]Quantum Technology Group, 2nd Institute of Physics, RWTH Aachen University, Aachen, 52074, Germany

**Abstract.** We introduce a novel optical interface between a singlet-triplet spin qubit and a photonic qubit which would offer new prospects for future quantum communication applications. The interface is based on a 220 nm thick GaAs/Al-GaAs heterostructure membrane and features a gate-defined singlet-triplet qubit, a gate-defined optically active quantum dot, a photonic crystal cavity and a bottom gold reflector. All essential components can be lithographically defined and deterministically fabricated, which greatly increases the scalability of on-chip integration. According to our FDTD simulations, the interface provides an overall coupling efficiency of 28.7% into a free space Gaussian beam, assuming an $SiO_2$ interlayer filling the space between the reflector and the membrane. The performance can be further increased to 48.5% by undercutting this $SiO_2$ interlayer below the photonic crystal.

**Keywords:** quantum technology, gate-defined quantum dot, optical interface, singlet-triplet spin qubit, photonic crystal cavity, Gaussian beam.

## 1 Introduction

Interconnecting distant stationary qubits with photonic qubits is essential for advancing the quantum Internet [1], distributing quantum entanglement across various quantum technology platforms for applications like secure quantum communication [2-3] and distributed quantum computing [4]. Gate-defined quantum dots (GDQDs), which trap single electrons in tunable electrostatic potential minima within a two-dimensional electron gas (2DEG), emerge as promising candidates due to their compatibility with top-down fabrication techniques and the potential for integrating multiple locally interconnected qubits. Singlet-triplet qubits defined by GDQD based on gallium arsenide (GaAs) are particularly promising as they feature all necessary prerequisites for quantum information processing, including initialization, readout, all electrical control, and high-speed qubit-state manipulation [5-10]. The direct bandgap of GaAs further facilitates efficient photonic qubit to spin qubit conversion by direct optical absorption and emission.

However, achieving coherent and efficient interfacing between photonic and spin qubits in GDQD systems poses challenges, notably due to the lack of hole confinement in conventional heterostructures. A solution involves integrating an optically active quantum dot (OAQD), such as a fully gate-defined electrostatic exciton trap [11], as an intermediary for coherently transferring photogenerated electrons to GDQD spin qubits [12]. This approach, complemented by advancements in optical interface technologies such as photonic crystal cavities (PCCs), enables enhanced light-matter interaction [13-17] and efficient photon emission [18-20].

In this paper and in line with our previous publications [21-23], we introduce a novel design of a singlet-triplet spin qubit to fiber interface incorporating a gate-defined OAQD at the center of a PCC. All the structures are fully lithographically defined on a 220 nm GaAs/AlGaAs heterostructure membrane. According to our theoretical modeling and numerical evaluation, this design ensures that more than 50% of the photons emitted by the OAQD are coupled into a narrow free space beam in the perpendicular direction, which in turn has an optical overlap greater than 50% with an ideal Gaussian beam, facilitating coupling to an optical fiber. This integration approach is not only compatible with electrode integration into the photonic cavity, but also promises to significantly advance scalability by means of top-down fabrication. It thus provides a practical path towards the efficient and scalable integration of GDQD and photonic cavities for quantum information applications.

## 2     Modeling of the optical interface

Fig. 1(a) presents a top view of our optical interface, which consists of a PCC defined by a triangle lattice of holes (black circles), cavity openings with mini-stopband (light and dark blue circles) providing electrical connectivity to the surrounding 2DEG, an Au/Ti metal gate system defining the GDQD, the OAQD and a single electron transistor (SET) used to read out the state of the GDQD. The detailed gate structure is depicted in Fig. 1(b). The entire structure, including the PCC and the gate system, is lithographically defined on a 220 nm thick GaAs/AlGaAs membrane (Fig. 1(c)). The OAQD emission wavelength is 823 nm at cryogenic temperatures and can be tuned from 820 nm to 825 nm by adjusting the electrical voltage applied to the trap and the guard gate (quantum-confined Stark effect [24]). A 200 nm thick gold reflector (not shown) is attached to the bottom side of the membrane via an $SiO_2$ interlayer to recycle photons back to the desired positive z-direction.

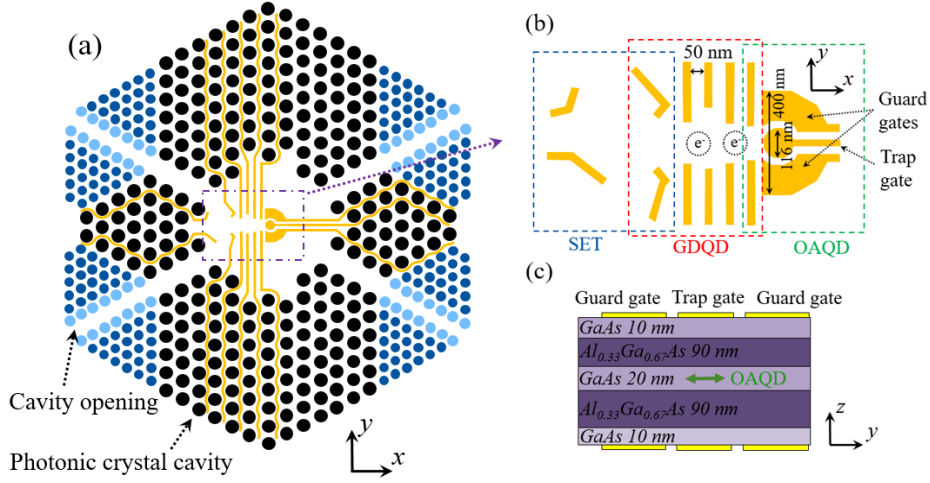

**Fig. 1.** (a) Top view of the optical interface including the PCC, the electrode system, and the cavity openings. (b) Detailed gate structure of the SET, the GDQD and the OAQD. The Au/Ti wires have a thickness of 9 nm and a width of 30 nm. Gates for the OAQD are fabricated on both sides of the membrane, while the SET and the GDQD only require top side electrodes. (c) Cross sectional view of the OAQD in the *yz*-plane through the center of the trap gate.

The PCC is created by modifying a hexagonal H4 cavity [21], with a lattice constant $a_1$ = 290 nm and a hole radius $r_1$ = 109 nm (lattice of black circles). When such a lattice is etched in the 220 nm GaAs/AlGaAs heterostructure membrane (Fig. 1(c)), a transverse electric (TE) photonic bandgap is opened from 717 nm to 1021 nm.

The four cavity openings in the diagonal directions are created by removing a row of etched holes. Those cavity openings are required to ensure a continuous 2DEG in the GaAs quantum well layer located at the center of the stack, which is an essential prerequisite for the functionality of the SET and of the GDQD. The underlying reason is that the surface charges trapped on the side walls of the etched holes deplete the free carriers in their vicinity, which prohibits electron transport through the photonic crystal lattice except in the cavity openings.

On the other hand, to maintain optical confinement inside the PCC, we modified the lattice constant and the hole radii of the PCC near the cavity openings. As shown in Fig. 1(a), the lattice marked in light and dark blue circles has a reduced lattice constant of $a_2$ = 214 nm (down from $a_1$ = 290 nm), and reduced radii of $r_2$ = 90 nm and $r_3$ = 73 nm, respectively (down from $r_1$ = 109 nm). These cavity openings support a TE mini-stopband [25] opening from 793 nm to 845 nm, i.e., no propagating mode is allowed to transmit though the cavity openings within this wavelength range.

## 3    Simulation results

To evaluate the theoretical performance of the complete structure, we performed three-dimensional finite-difference time-domain (3D FDTD) simulations. All the structures shown in Fig. 1 are reproduced in the simulations (Lumerical FDTD). The OAQD is modeled by a dipole orientated in the *y*-direction located at the center of the PCC. Modeling the dipole this way is justified by considering that the in-plane polarization of the emitted photon is determined by the external magnetic field and by the heavy-hole nature of the participating hole [12]. The refractive indices of GaAs and $Al_{0.33}Ga_{0.67}As$ utilized in the simulations are approximated to be $n_{4K, GaAs}$ = 3.59 and $n_{4K, AlGaAs}$ = 3.38 by considering their temperature dependence and linearly extrapolating the room temperature values [26]. The absorption coefficients of GaAs and AlGaAs are assumed to be zero in the simulations. The thickness of the $SiO_2$ interlayer is optimized in such a way that a high extraction efficiency is achieved at the working wavelength (823 nm) of the OAQD.

Fig. 2 shows the simulated cavity spectrum from 790 nm to 870 nm excited by a y-dipole located in the center of the PCC. Due to the relatively large size of the PCC, multiple resonant peaks are observed in this wavelength range. The target mode at 823 nm, whose wavelength coincides with the emission wavelength of the OAQD as a result of the optimization of the PCC parameters, is highlighted with a green arrow. For comparison to the target mode, another mode at 833 nm is also highlighted by a brown arrow.

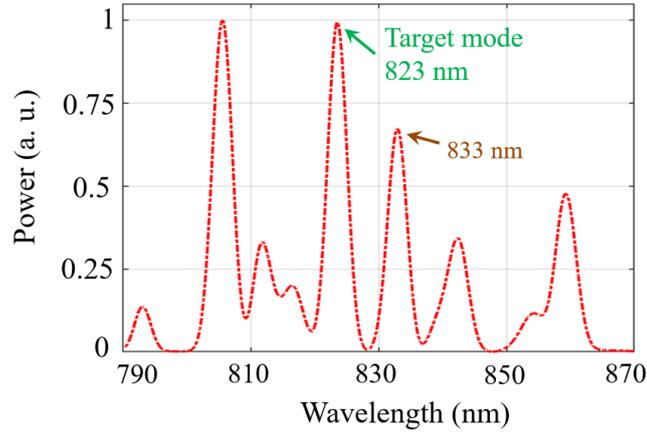

**Fig. 2.** Cavity power spectrum in the PCC. The target mode, which has a high extraction efficiency, is marked by a green arrow at 823 nm. For comparison, another cavity mode at 833 nm is also highlighted by a brown arrow.

To study the far-field emission properties, we record the electromagnetic field with a 5 μm by 5 μm 2D monitor located 20 nm above the top surface of the membrane and apply a far-field transform to the recorded near-field. The far-field patterns at a distance 1 m away from the membrane for both cavity modes are depicted in Fig. 3. It is clear that the target cavity mode at 823 nm features a pronounced single-beam vertical emission at x = 0 and y = 0. On the other hand, the far-field emission pattern of the cavity mode at 833 nm shows a complex profile not easily matched to that of an optical fiber.

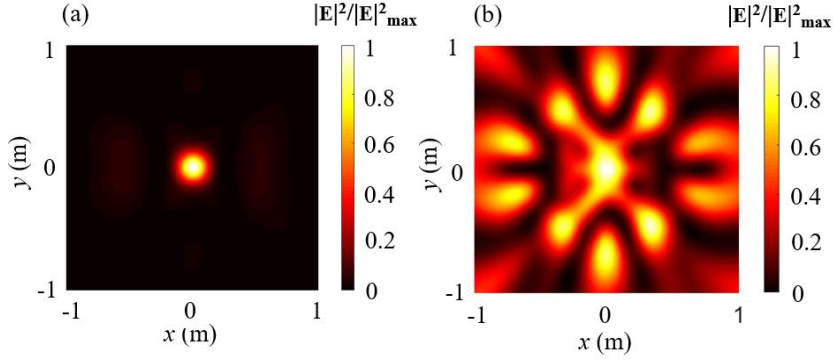

**Fig. 3.** Far-field patterns of the two highlighted cavity modes at (a) 823 nm and (b) 833 nm, at a distance 1 m away from the membrane. The intensities are independently normalized by the maximum value for both far-field patterns.

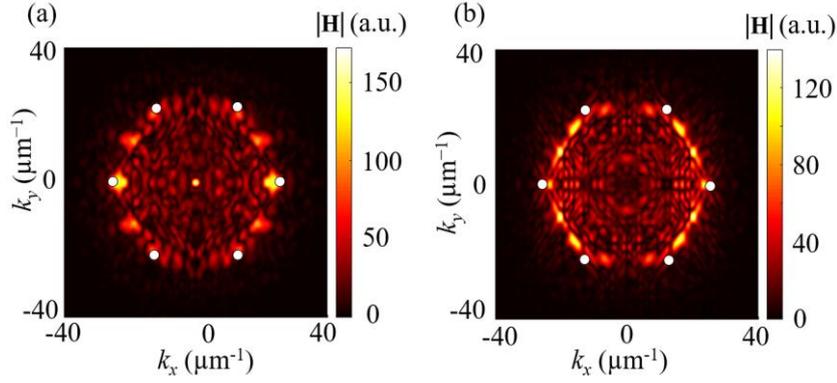

**Fig. 4.** Fourier space distribution of the cavity mode at (a) 823 nm and (b) 833 nm. The white dots represent the reciprocal lattice points of the photonic crystal lattice with a lattice pitch 290 nm.

The difference in the far-field emission pattern can be explained by applying a spatial 2D Fourier transformation to the recorded near-field electromagnetic waves and comparing the dominant k-space components with the reciprocal lattice points of the photonic crystal (black lattice in Fig. 1). For the target mode, we observe that the cavity mode features dominant $k$-space components at the reciprocal lattice points of the photonic crystal (Fig. 4(a)). As a result, these components experience a Bragg-scattering to the $\Gamma$-point ($k_x = k_y = 0$) and result in a pronounced vertical emission. On the other hand, the dominant k-space components of the cavity mode at 833 nm do not overlap with the reciprocal lattice points (Fig 4(b)), which explains the complex far-field emission pattern seen in Fig. 3(b).

To rigorously evaluate the probability that a photon emitted by the OAQD couples into the fundamental mode of a single mode fiber, we first calculate the possibility of the emitted photon escaping the semiconductor membrane and being recorded by the monitor (escape probability). This is done by integrating the normal component of the Poynting vector over the entire monitor surface and dividing the integrated power by the total power emitted by the dipole. Then, we calculate the vectorial optical overlap (OV) between the recorded near-field and an ideal 2D Gaussian beam according to [27]

$$\mathrm{OV}(\theta) = \frac{|\iint (\vec{E}_m \times \vec{H}_g^*(\theta) + \vec{E}_g^*(\theta) \times \vec{H}_m) \cdot d\vec{S}|^2}{4 \iint Re(\vec{E}_m \times \vec{H}_m^*) \cdot d\vec{S} \iint Re(\vec{E}_g(\theta) \times \vec{H}_g^*(\theta)) \cdot d\vec{S}} \quad (1)$$

where $\vec{E}_m$ ($\vec{H}_m$) is the electric (magnetic) near-field recorded by the monitor, and $\vec{E}_g(\theta)$ ($\vec{H}_g(\theta)$) represents the electric (magnetic) field of a linearly polarized Gaussian beam with an $1/e^2$ intensity half angle $\theta$.

The multiplication of the escape probability with OV($\theta$) gives the overall efficiency, which is equal to the overall probability of the photon emitted by the dipole to couple to a Gaussian beam with a half angle $\theta$. Fig. 5 shows the overall efficiency and the OV as a function of $\theta$ for both highlighted cavity modes.

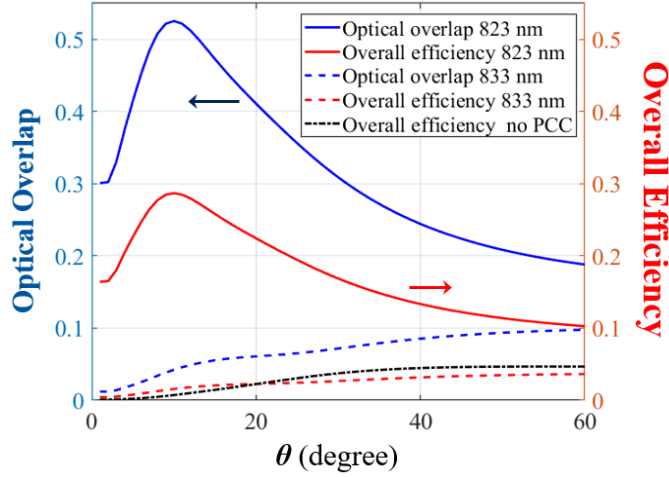

**Fig. 5.** OV and overall efficiency as a function of $\theta$ for both cavity modes. To show the effectiveness of the PCC, we also plot the overall efficiency when the PCC is removed (black dotted curve).

As we can see in Fig. 5, the maximum OV and overall efficiency for the target cavity mode are 0.526 and 28.7%, respectively, at a divergence angle of 10 degrees with an escape probability of 54.6%. In comparison, the OV for the 833 nm cavity mode stays below 0.1, and the overall efficiency for the 833 nm cavity mode and for an electrode system without the PCC stays below 5% for all $\theta$. In addition, we also investigated the performance when the SiO$_2$ interlayer is removed and the PCC fabricated in a suspended membrane. After re-optimizing the position of the bottom reflector, the maximum OV reaches 0.839 with a maximum overall efficiency of 48.5% at a divergence angle of 10 degrees. While this configuration increases the fabrication complexity, it offers a path forward for a significant improvement of the device performance.

## 4 Conclusion

In this paper, we present an efficient optical interface, assisted by a photonic crystal cavity, between a singlet-triplet spin qubit and a photonic qubit. The entire structure can be lithographically defined and fabricated. By utilizing the Bragg scattering of the PCC, we obtain vertical Gaussian emission at the desired wavelength. According to our calculations, our design reaches an overall efficiency of 28.7% by emitting into a Gaussian beam with a divergence angle of 10 degrees. The performance can be further increased by removing the $SiO_2$ interlayer. In this case, the overall efficiency is 48.5% at the same divergence angle. Fabrication of this device is currently in progress and we expect to be able to report first experimental results in the near future.

## 5 Acknowledgement

This project is funded by the Deutsche Forschungsgemeinschaft (DFG, German Research Foundation) under Germany's Excellence Strategy – Cluster of Excellence Matter and Light for Quantum Computing (ML4Q) EXC 2004/1 – 390534769.